\documentclass[10pt,amsmath,aps,pre,twocolumn]{revtex4-1}

\usepackage{graphicx,hyperref}
\usepackage[varg]{txfonts}

\newcommand{\ii}{\mathrm{i}}
\newcommand{\e}{\mathrm{e}}
\renewcommand{\d}{\mathrm{d}}

\newcommand{\tens}[1]{\mathbf{#1}}
\newcommand{\cvector}[1]{\left(\begin{array}{c}#1\end{array}\right)}

\newcommand{\mnras}{MNRAS}
\newcommand{\physrep}{Phys.~Rep.}

\begin{document}

\title{Non-equilibrium statistical field theory for classical particles: Non-linear structure evolution with first-order interaction}
\author{Matthias Bartelmann, Felix Fabis, Daniel Berg, Elena Kozlikin, Robert Lilow, Celia Viermann}
\affiliation{Heidelberg University, Zentrum f\"ur Astronomie, Institut f\"ur Theoretische Astrophysik, Philosophenweg 12, 69120 Heidelberg, Germany}

\begin{abstract}
We calculate the power spectrum of density fluctuations in the statistical non-equilibrium field theory for classical, microscopic degrees of freedom to first order in the interaction potential. We specialise our result to cosmology by choosing appropriate initial conditions and propagators and show that the non-linear growth of the density power spectrum found in numerical simulations of cosmic structure evolution is reproduced well to redshift zero and for arbitrary wave numbers. The main difference of our approach to ordinary cosmological perturbation theory is that we do not perturb a dynamical equation for the density contrast. Rather, we transport the initial phase-space distribution of a canonical particle ensemble forward in time and extract any collective information from it at the time needed. Since even small perturbations of particle trajectories can lead to large fluctuations in density, our approach allows to reach high density contrast already at first order in the perturbations of the particle trajectories. We argue why the expected asymptotic behaviour of the non-linear power spectrum at large wave numbers can be reproduced in our approach at any order of the perturbation series.
\end{abstract}

\maketitle

\section{Introduction}

In pioneering papers, \citeauthor{2010PhRvE..81f1102M} and \citeauthor{2012JSP...149..643D} have reformulated kinetic theory as a non-equilibrium, statistical field theory for classical particles \cite{2010PhRvE..81f1102M, 2011PhRvE..83d1125M, 2012JSP...149..643D, 2013JSP...152..159D}. Based on their work, we have derived initial conditions for canonical ensembles of Hamiltonian point particles in phase space \cite{2014arXiv1411.0806B} and showed how the power spectrum and the bispectrum of the density contrast of linearly and mildly non-linearly evolved cosmic structures could be derived from this theory \cite{2014arXiv1411.1153B}. Here, we proceed to calculate the non-linear evolution of the cosmological density power spectrum to first order in the interaction potential between the point particles.

The essential difference between our approach and the formidable body of work on cosmological perturbation theory (see \cite{1980lssu.book.....P, 2002PhR...367....1B} for reviews) is that we do not derive, use or perturb any dynamical equation for the density contrast itself. Rather, we describe canonical ensembles of microscopic particles in cosmology by initial conditions suitably correlated in phase space, which are propagated forward in time by the Green's function of the free Hamiltonian (see also \cite{2014arXiv1411.0805B}). The initial conditions and the evolution are encoded in a free generating functional \cite{2014arXiv1411.0806B}.

Any collective information on the ensemble, most notably on the matter density field composed of the particles, is embodied by a collective-field operator multiplied to the free generating functional. In close analogy to statistical quantum field theory, the interaction between the point particles is represented by another exponential operator acting on the free generating functional. Taylor-expanding this interaction operator leads to the Feynman diagrams of quantum field theory. We follow the same approach here, expand the interaction operator to first order in the interaction potential, and calculate the evolution of the density power spectrum to this order.

Thus, it is not the density contrast whose evolution we study in a perturbative manner. Instead, we read off the density and its second-order cumulant from the generating functional of this non-equilibrium statistical theory at the time needed. The evolution of the canonical particle ensemble is described by the retarded Green's function of the particle trajectories in phase space. Since even weak perturbations of trajectories can lead to strong perturbations of the density, one decisive advantage of this approach is that we can proceed deeply into the regime of non-linear density perturbations.

We summarise the first-order perturbative approach in our theory in Sect.~2 and proceed to calculate third- and fourth-order density cumulants in Sect.~3. The non-linear evolution of the cosmic-density power spectrum in this first-order perturbative approach is calculated in Sect.~4, and we summarise our conclusions in Sect.~5. Clearly, first-order perturbation theory cannot be expected to produce the final answer. Furthermore, we are combining two different types of propagators and treat the damping factor inevitably appearing in our theory in a rather approximate manner.

Nonetheless, our results seem to show that, even at low perturbative order, the non-linear power spectrum of cosmic density perturbations can be calculated with our theory in an analytic and rather simple way. Our approximation to the non-linear cosmic power spectrum is valid to redshift zero, extends to arbitrary wave numbers, and has no free parameters.

\section{First-order perturbation theory in the canonical ensemble}

\subsection{One- and two-point cumulants with first-order interaction}

We have shown in \cite[Eq.~65]{2014arXiv1411.0806B} that the generating functional including interaction can be created from the free generating functional $Z_0[\tens J, \tens K]$ by means of an interaction operator,
\begin{equation}
  Z[H,\tens J,\tens K] = \e^{\ii\hat S_\mathrm{I}}Z_0[H,\tens J,\tens K] = 
  \e^{\ii\hat S_\mathrm{I}}\e^{\ii H\cdot\hat\Phi}Z_0[\tens J,\tens K]\;,
\label{eq:04-1}
\end{equation}
with the interaction part of the action given by the operator
\begin{equation}
  \hat S_\mathrm{I} = \int\d 1\int\d 2\left(
    \frac{\delta}{\delta H_B(2)}v(12)\frac{\delta}{\delta H_\rho(1)}
  \right)\;,
\label{eq:04-2}
\end{equation}
defined with slightly more explicit notation in Eq.~(64) of \cite{2014arXiv1411.0806B}. Here, the interaction potential between the two positions $1$ and $2$ is
\begin{equation}
  v(12) := v\left(\vec q_1-\vec q_2\right)
  \delta_\mathrm{D}\left(\tau_1-\tau_2\right)\;.
\label{eq:04-3}
\end{equation} 
The time $\tau$ generalises the coordinate time $t$ here. Equation (\ref{eq:04-3}) contains two assumptions on the potential which will become important shortly. First, it is assumed to be translation invariant and thus depends on the coordinate difference $\vec q_1-\vec q_2$ only. Second, it is assumed to act instantaneously, expressed by the delta distribution in time.

Since the functional derivatives with respect to $H$ act only on the collective-field operator $\e^{\ii H\cdot\hat\Phi}$, the effect of the interaction operator can be brought into the form
\begin{equation}
  Z[H,\tens J,\tens K] = \e^{\ii H\cdot\hat\Phi}
  \e^{\ii S_\mathrm{I}}Z_0[\tens J,\tens K]
\label{eq:04-4}
\end{equation}
with
\begin{equation}
  S_\mathrm{I} = -\int\d 1\int\d 2\,
    \hat\Phi_B(2)\,v(12)\,\hat\Phi_\rho(1)\;.
\label{eq:04-5}
\end{equation}
The density and response-field operators, $\hat\Phi_\rho$ and $\hat\Phi_B$, in the interaction part $S_\mathrm{I}$ of the action now act directly on the free generating functional and produce cumulants of the form studied in \cite{2014arXiv1411.1153B}. To lowest non-trivial order, the interaction operator is
\begin{equation}
  \e^{\ii S_\mathrm{I}} \approx 1-
  \ii\int\d 1\int\d 2\hat\Phi_B(2)\,v(12)\,\hat\Phi_\rho(1)\;.
\label{eq:04-6}
\end{equation}

The corrections to the one- and two-point density cumulants in first non-trivial order are then
\begin{align}
  \delta^{(1)}G_\rho(1) &=
  \hat\Phi_\rho(1)\left(-\ii S_\mathrm{I}Z_0[\tens J,\tens K]\right)
  \nonumber\\ &=
  -\ii\int\d1'\int\d2'\,v(1'2')\,G_{B\rho\rho}(11'2')
\label{eq:04-7}
\end{align}
and similarly
\begin{equation}
  \delta^{(1)}G_{\rho\rho}(12) =
  -\ii\int\d 1'\int\d 2'\,v(1'2')\,G_{B\rho\rho\rho}(121'2')\;.
\label{eq:04-8}
\end{equation}
Note that we now denote with primes the internal vertices of the interaction, which are integrated over in the interaction operator.

As we have seen in \cite[Eqs.~51 and 52]{2014arXiv1411.0805B}, a one-particle response-field operator $\hat\Phi_{B_{j_m}}(m)$ acting on the free generating functional following $(m-1)$ one-particle density operators results in
\begin{align}
  &\hat\Phi_{B_{j_m}}(m)\left.
    \left(\hat\Phi_{\rho_{j_{m-1}}}(m-1)\ldots\hat\Phi_{\rho_{j_1}}(1)\right)
  \right\vert_{\tens J=0}Z_0[\tens J, \tens K] \nonumber\\ &=
  b_{j_m}(m)\left.
    \left(\hat\Phi_{\rho_{j_m}}(m)\ldots\hat\Phi_{\rho_{j_1}}(1)\right)
  \right\vert_{\tens J=0}Z_0[\tens J, \tens K] \nonumber\\ &=
  b_{j_m}(m)G_{\rho_{j_m}\ldots\rho_{j_1}}(1\ldots m)
\label{eq:04-9}
\end{align}
with the one-particle response-field factor
\begin{equation}
  b_{j_m}(m) = \ii\sum_{s=1}^mg_{qp}(t_s,t_m)
  \vec k_m\cdot\vec k_s\,\delta_{j_mj_s}\;.
\label{eq:04-10}
\end{equation}
It will be important for later calculations to note here that the Kronecker delta in the response-field factor couples two particles.

For calculating the first-order approximation of the non-linear density evolution and the non-linear power spectrum, we thus have to work out the three- and four-point cumulants $G_{B\rho\rho}(11'2')$ and $G_{B\rho\rho\rho}(121'2')$ of the free generating functional.

Before doing so, we notice that the integrals in (\ref{eq:04-7}) and (\ref{eq:04-8}) need to be carried out in configuration space, while the three- and four-point cumulants will be given in Fourier space. For the following brief calculation at the example of the three-point cumulant, we denote with $\tilde G_{B\rho\rho}(11'2')$ the cumulant in configuration space rather than in real space.

We write the spatial part of the integral (\ref{eq:04-7}) as
\begin{align}
  &\int\d^3q_1'\int\d^3q_2'\,v(1'2')\,\tilde G_{B\rho\rho}(11'2')
  \nonumber\\ &=
  \int\d^3q_1'\int\d^3q_2'\,v(1'2') \nonumber\\ &\times
  \int\frac{\d^3k_1'}{(2\pi)^3}\int\frac{\d^3k_2'}{(2\pi)^3}
  G_{B\rho\rho}(11'2')
    \e^{\ii\vec k_1'\cdot\vec q_1'+\ii\vec k_2'\cdot\vec q_2'}
  \nonumber\\ &=
  \int\frac{\d^3k_1'}{(2\pi)^3}
  \hat v\left(\vec k_1'\right)G_{B\rho\rho}(11'{-1'})\;.
\label{eq:04-11}
\end{align}
We have used here that the potential is assumed to be translation invariant, which introduces a delta distribution $\delta_\mathrm{D}(\vec k_1'+\vec k_2')$ replacing the argument $2'$ by $-1'$ in the last step.

Consequently, as we shall work out the three- and four-point cumulants $G_{B\rho\rho}(11'2')$ and $G_{B\rho\rho\rho}(121'2')$, we shall be allowed to simplify terms by setting $\vec k_1'+\vec k_2' = 0$. Since the potential is additionally assumed to act instantaneously, we may set $\tau_1' = \tau_2'$.

Moreover, if we can assume that the potential depends on the modulus of its argument only, its Fourier transform must be real,
\begin{align}
  \hat v\left(\vec k\,\right) =
  4\pi\int_0^\infty x\d x\,v(x)\frac{\sin kx}{k} \in\mathbb{R}\;.
\label{eq:04-12}
\end{align}
Since the potential is also real in configuration space, this implies that its Fourier transform is symmetric,
\begin{equation}
  \hat v\left(\vec k\,\right) = \hat v^*\left(-\vec k\,\right) =
  \hat v\left(-\vec k\,\right)\;.
\label{eq:04-13}
\end{equation} 

For calculating the first-order effects of the interaction potential on the mean density and the density power spectrum, we thus have to work out the two expressions
\begin{equation}
  \delta^{(1)}G_\rho(1) =
  -\ii\int_0^{\tau_1}\d\tau_1'\int\frac{\d^3k_1'}{(2\pi)^3}
  \hat v\left(\vec k_1^{\,\prime}\right)G_{B\rho\rho}(11'{-1'})
\label{eq:04-14}
\end{equation}
and
\begin{equation}
  \delta^{(1)}G_{\rho\rho}(12) =
  -\ii\int_0^{\tau_1}\d\tau_1'\int\frac{\d^3k_1'}{(2\pi)^3}
  \hat v\left(\vec k_1^{\,\prime}\right)G_{B\rho\rho\rho}(121'{-1'})\;,
\label{eq:04-15}
\end{equation} 
both to be evaluated at $\tau_1' = \tau_2'$.

\subsection{Cumulants}

As described in \cite{2014arXiv1411.1153B}, the density cumulants are conveniently decomposed into their one-particle contributions,
\begin{equation}
  G_{\rho\ldots\rho}(1\ldots m) = \sum_{j_1\ldots j_m=1}^NG_{j_1\ldots j_m}\;,
\label{eq:04-16}
\end{equation}
which are determined by the free generating functional $\bar Z_0[\tens L]$ evaluated at the shift $\tens L$,
\begin{equation}
  G_{j_1\ldots j_m} = \bar Z_0[\tens L]\;,
\label{eq:04-17}
\end{equation}
created by the density operators,
\begin{equation}
  \tens L(\tau) = -\sum_{s=1}^m\delta_\mathrm{D}\left(\tau-\tau_s\right)
  \cvector{\vec k_s\\ 0}\otimes\vec e_{j_s}\;.
\label{eq:04-18}
\end{equation}
Its time-integrated components $\tens L_q$ and $\tens L_p$ have the components
\begin{equation}
  \bar L_{q_j} = -\sum_{s=1}^m\vec k_s\delta_{jj_s}\;,\quad
  \bar L_{p_j} = -\sum_{s=1}^mg_{qp}(\tau_s,0)\vec k_s\delta_{jj_s}\;.
\label{eq:04-18a}
\end{equation}
For the sake of a more compact notation, we abbreviate
\begin{equation}
  \vec K_s := g_{qp}(\tau_s,0)\vec k_s
\label{eq:04-18b}
\end{equation}
in the following.

Regarding the generating functional $\bar Z_0[\tens L]$ itself, we have shown in \cite{2014arXiv1411.0806B} that it can be approximated by
\begin{equation}
  \bar Z_0[\tens L] \approx \bar Z_0^{(1)}[\tens L]+\bar Z_0^{(2)}[\tens L]\;,
\label{eq:04-19}
\end{equation}
with the terms on the right-hand side being due to linear and quadratic momentum correlations, respectively.

These terms are sums over contributions by individual pairs of different positions,
\begin{align}
  \bar Z_0^{(1)}[\tens L] &= V^{-N}\e^{-Q_D/2}\sum_{j\ne k=1}^N
  \bar Z_{jk}^{(1)} \;,\nonumber\\
  \bar Z_0^{(2)}[\tens L] &= \frac{V^{-N}}{8}
  \e^{-Q_D/2}\sum_{j\ne k, l\ne m=1}^N\bar Z_{jklm}^{(2)}
\label{eq:04-20}
\end{align}
with
\begin{align}
  \bar Z_{jk}^{(1)} = 
  (2\pi)^3\delta_\mathrm{D}\left(\bar L_{q_j}+\bar L_{q_k}\right)
  \mathcal{N}'_{jk}
  P_\delta\left(\bar L_{q_j}\right)A_{jk}^2\left(\bar L_{q_j}\right)\;.
\label{eq:04-21}
\end{align}
Of the quadratic terms $\bar Z_{jklm}^{(2)}[\tens L]$ derived in \cite{2014arXiv1411.0806B}, we shall here only need the two-point contribution
\begin{align}
  &\bar Z_{jkjk}^{(2)} = (2\pi)^3
  \delta_\mathrm{D}\left(\bar L_{q_j}+\bar L_{q_k}\right)\mathcal{N}'_{jk}
  \nonumber\\ &\times
  \int\frac{\d^3k}{(2\pi)^3}
  P_\delta\left(\vec k\,\right)P_\delta\left(\vec k-\bar L_{q_j}\right)
  a_{jk}^2\left(\vec k\,\right)a_{jk}^2\left(\vec k-\bar L_{q_j}\right)
\label{eq:04-22}
\end{align}
and the three-point contribution
\begin{align}
  &\bar Z_{jkkl}^{(2)} = (2\pi)^3
  \delta_\mathrm{D}\left(\bar L_{q_j}+\bar L_{q_k}+\bar L_{q_l}\right)
  \mathcal{N}'_{jkl} \nonumber\\ &\times
  P_\delta\left(\bar L_{q_j}\right)P_\delta\left(\bar L_{q_l}\right)
  a_{jk}^2\left(\bar L_{q_j}\right)a_{kl}^2\left(\bar L_{q_l}\right)\;.
\label{eq:04-23}
\end{align}

In the preceding terms, the abbreviation
\begin{equation}
  \mathcal{N}'_{jk} := \int\d\tens q'\,
  \e^{\ii\left\langle\bar{\tens L}_q,\tens q'\right\rangle}
\label{eq:04-24}
\end{equation}
was introduced, where prime indicates that the positions $\vec q_j$ and $\vec q_k$ are to be excluded from the integration over all spatial particle positions $\tens q$. Moreover,
\begin{equation}
  A_{jk}^2\left(\bar L_{q_j}\right) := \frac{1}{2}\left(
    1-a_{jk}^2\left(\bar L_{q_j}\right)
  \right)-b_{jk}\left(\bar L_{q_j}\right)
\label{eq:04-25}
\end{equation}
was defined, containing
\begin{equation}
  a_{jk}^2\left(\bar L_{q_j}\right) :=
  \frac{\left(\bar L_{p_j}\cdot\bar L_{q_j}\right)
        \left(\bar L_{q_j}\cdot\bar L_{p_k}\right)}
  {\bar L_{q_j}^{\,4}} \;,\quad
  b_{jk}\left(\bar L_{q_j}\right) :=
  \frac{\bar L_{q_j}\cdot\bar L_{p_k}}{\bar L_{q_j}^{\,2}}\;.
\label{eq:04-26}
\end{equation}
The terms $\bar Z_{jk}^{(1)}$ are not necessarily symmetric in $(j,k)$ because of the $B_{xp}$ correlation between densities and momenta. By construction, the terms $\bar Z_{jklm}^{(2)}$ are symmetric under the permutations $(jklm)\to(lmjk)$, $(jklm)\to (kjlm)$ and $(jklm)\to(jkml)$.

\subsection{Gravitational particle interaction in cosmology}

The Lagrange function of a point particle with mass $m$ in a homogeneously and isotropically expanding space-time is
\begin{equation}
  L\left(\vec q,\dot{\vec q},t\right) =
  \frac{m}{2}a^2\dot{\vec q}^{\,2}-m\phi\;,
\label{eq:04-27}
\end{equation}
where $\vec q$ is the particle position in comoving coordinates, and $\phi$ satisfies the Poisson equation
\begin{equation}
  \vec\nabla_q^2\phi = 4\pi Ga^2m(\rho-\bar\rho)\;,
\label{eq:04-28}
\end{equation}
sourced by the fluctuation $(\rho-\bar\rho)$ of the \emph{number} density $\rho$ about its mean $\bar\rho$ \cite{1980lssu.book.....P}. We now transform the time from the cosmic time $t$ to the new time coordinate
\begin{equation}
  \tau := D_+(t)-1\;,
\label{eq:04-29}
\end{equation}
where $D_+(t)$ is the linear growth factor of cosmic density fluctuations normalised to unity at the initial time. If we further pull a factor $mH_\mathrm{i}$ out of the Lagrangian, $H_\mathrm{i}$ being the Hubble function at the initial time $\tau = 0$, the Lagrange function transforms into
\begin{equation}
  L\left(\vec q,\dot{\vec q},\tau\right) =
  \frac{g(\tau)}{2}\dot{\vec q}^{\,2}-v\left(\vec q,\tau\right)\;,
\label{eq:04-30}
\end{equation}
with the potential
\begin{equation}
  v = \frac{a^2\phi}{g(\tau)H_\mathrm{i}^2}\;;
\label{eq:04-31}
\end{equation}
cf.\ \cite{2014arXiv1411.0805B}. Like the growth factor $D_+$, the cosmological scale factor $a$ is normalised to unity at the initial time $\tau = 0$. The function $g(\tau)$ is defined as
\begin{equation}
  g(\tau) := a^2D_+fHH_\mathrm{i}^{-1}
\label{eq:04-32}
\end{equation}
with
\begin{equation}
  f := \frac{\d\ln D_+}{\d\ln a}\;.
\label{eq:04-33}
\end{equation}
For an Einstein-de Sitter universe, $D_+ = a$, $f = 1$ and $g(\tau) = a^{3/2} = (1+\tau)^{3/2}$.

Introducing the density contrast $\delta$ and the mean cosmic particle number density $\bar\rho$,
\begin{equation}
  \delta := \frac{\rho-\bar\rho}{\bar\rho}\;,\quad
  \bar\rho = \frac{3H_\mathrm{i}^2}{8\pi Gm}\Omega_\mathrm{m,i}a^{-3}
\label{eq:04-34}
\end{equation}
we see that the potential $v$ needs to satisfy the Poisson equation
\begin{equation}
  \vec\nabla_q^2v = \frac{3}{2}\frac{a}{g(\tau)}\delta\;.
\label{eq:04-35}
\end{equation}
We now write the density contrast as
\begin{equation}
  \delta = \bar\rho^{-1}
  \sum_{j=1}^N\delta_\mathrm{D}\left(\vec q-\vec q_j\right)-1\;,
\label{eq:04-36}
\end{equation}
Fourier transform the Poisson equation and consider the contribution from a single particle at the coordinate origin. Then, the Fourier transform of the potential of a single particle is
\begin{equation}
  \hat v\left(\vec k\,\right) =
  -\frac{3}{2}\frac{a}{g(\tau)}\left(\frac{1}{\bar\rho k^2}-\hat 1\right)\;.
\label{eq:04-37}
\end{equation}
The Fourier-transformed unity $\hat 1$ can be neglected later because the zero mode will not contribute to any cumulants. We can thus insert
\begin{equation}
  v\left(\vec k\,\right) = -\frac{3}{2}\frac{a}{g(\tau)}\frac{1}{\bar\rho k^2}
\label{eq:04-38}
\end{equation}
for the Fourier-transformed, one-particle potential. Notice in particular that this potential scales inversely with the mean particle density $\bar\rho$. This is because, for a fixed mean mass per volume, the particle mass has to decrease in inverse proportion to the particle number $N$ if that number is increased.

\subsection{Shot noise and the relevance of terms}

In our microscopic approach, shot-noise terms appear because the density field is composed of discrete particles. To identify these terms and to clarify their relevance, consider a statistically homogeneous density field
\begin{equation}
  \rho\left(\vec q\,\right) = \sum_{i=1}^N\delta_\mathrm{D}\left(\vec q-\vec q_i\right)
\label{eq:04-39}
\end{equation}
composed of $N$ point particles. In Fourier space, this density field is
\begin{equation}
  \hat\rho\left(\vec k\,\right) = \sum_{i=1}^N\e^{-\ii\vec k\cdot\vec q_i}\;.
\label{eq:04-40}
\end{equation}
In terms of the density contrast $\delta$, the power spectrum of the density field is
\begin{align}
  \left\langle
    \hat\rho\left(\vec k\,\right)\hat\rho\left(\vec k^{\,\prime}\right)
  \right\rangle &=
  \bar\rho^2\left(
    1+\left\langle
      \hat\delta\left(\vec k\,\right)\hat\delta\left(\vec k^{\,\prime}\right)
    \right\rangle
  \right) \nonumber\\ &=
  \bar\rho^2\left(
    1+(2\pi)^3\delta_\mathrm{D}\left(\vec k+\vec k'\right)
    P_\delta\left(\vec k\,\right)
  \right)
\label{eq:04-41}
\end{align}
by definition of the density-contrast power spectrum $P_\delta(\vec k\,)$. If the density fluctuations are uncorrelated,
\begin{equation}
  \left\langle\hat\rho\hat\rho'\right\rangle = \bar\rho^2\;.
\label{eq:04-42}
\end{equation} 

On the other hand, inserting (\ref{eq:04-40}) into (\ref{eq:04-42}) results in
\begin{align}
  &\left\langle\hat\rho\hat\rho'\right\rangle =
  \left\langle\sum_{i,j=1}^N\e^{-\ii\vec k\cdot\vec q_j-\ii\vec k^{\,\prime}\cdot\vec q_j}\right\rangle \nonumber\\ &=
  \left(\prod_{k=1}^N\int\frac{\d^3q_k}{V}\right)\left(
    \sum_{i=j=1}^N\e^{-\ii(\vec k+\vec k^{\,\prime})\cdot\vec q_i}+
    \sum_{i\ne j=1}^N\e^{-\ii\vec k\cdot\vec q_j-\ii\vec k^{\,\prime}\cdot\vec q_j}
  \right) \nonumber\\ &=
  \frac{N}{V}(2\pi)^3\delta_\mathrm{D}\left(\vec k+\vec k^{\,\prime}\right)+
  \frac{N(N-1)}{V^2}(2\pi)^3\delta_\mathrm{D}\left(\vec k\,\right)(2\pi)^3\delta_\mathrm{D}\left(\vec k^{\,\prime}\right) \nonumber\\ &=
  \bar\rho(2\pi)^3\delta_\mathrm{D}\left(\vec k+\vec k^{\,\prime}\right)+\bar\rho^2\hat 1^2\;,
\label{eq:04-43}
\end{align}
abbreviating the Fourier-transformed unity by $\hat 1$. Obviously, only the second term in (\ref{eq:04-43}) corresponds to the result (\ref{eq:04-42}) for the continuous density field, while the first arises only because the density field is composed of discrete particles. Thus, the first term in (\ref{eq:04-43}) is a shot-noise term which arises from summing over pairs of identical particles, as the calculation shows.

More generally, for $m$-point cumulants of density fields composed of discrete particles, an analogous calculation shows that terms proportional to all powers of $\bar\rho$ occur, $\bar\rho^s$, with $1\le s\le m$. Only the term proportional to $\bar\rho^m$ is not a shot-noise term. It is the only term arising from summing over combinations of particles which are all different. Terms proportional to powers of $\bar\rho^s$ with $s<m$ are all shot-noise terms in the sense that they arise because of the discrete nature of the density field. In the thermodynamic limit $N\to\infty$, the shot-noise terms can be neglected relative to the dominant term proportional to $\bar\rho^m$.

In the case of gravitational interaction between the microscopic particles, the interaction potential scales with the particle mass. Resolving the density field into an increasing number of particles while keeping the \emph{mass} density constant, the particle mass must be decreased proportional to $N^{-1}$. This repeats the argument made following (\ref{eq:04-38}): The Poisson equation then implies that the gravitational interaction potential must scale inversely with the mean \emph{number} density of particles, i.e.\ like $\bar\rho^{-1}$.

According to (\ref{eq:04-5}), the interaction operator from the interaction part $S_\mathrm{I}$ of the action increases the order of the density $\rho$ and the response field $B$ in the free cumulants by one each and multiplies with a potential. As (\ref{eq:04-10}) shows, the response field couples two particles, as expressed by the Kronecker symbol $\delta_{j_mj_s}$ there. Comparing this with our earlier result on the origin of shot-noise terms, we see that the coupling of particles by the response field only selects shot-noise terms from the free density cumulants because the only non-shot noise term in the free density cumulants arises from combinations of different particles, for which $\delta_{j_mj_s} = 0$.

Specifically, for an $m$-point density cumulant in $n$-th order perturbation theory, free cumulants of order up to $m+2n$ need to be calculated which are of $(m+n)$-th order in the density and $n$-th order in the response field. In these free cumulants, terms proportional to all powers of $\bar\rho$ up to $\bar\rho^{m+2n}$ will occur. Their subsequent multiplication by $v^n$ will reduce the power of $\bar\rho$ by $n$ to $\bar\rho^{m+n}$. Each response field will couple particles pairwise and will thus further reduce the power of the leading term to $\bar\rho^m$, as expected for an $m$-point density cumulant.

This shows that only such terms in the free cumulants of order $m+2n$ need to be considered which are proportional to $\bar\rho^{m+n}$. Terms proportional to lower powers of $\bar\rho$ will vanish in the limit $N\gg1$, while terms proportional to higher powers of $\bar\rho$ disappear because of the coupling of particles by the response fields.

\section{Three- and four-point cumulants}

After these preparatory considerations, we shall now proceed to work out the three- and four-point cumulants $G_{B\rho\rho}(11'{-1'})$ and $G_{B\rho\rho\rho}(121'{-1'})$ we require. For all calculations carried out below, it is important that the response field couples two particles, which is mathematically expressed by the Kronecker delta in (\ref{eq:04-10}). Effectively, therefore, $m$-point cumulants of the form $G_{B\rho\ldots\rho}$ couple $m-1$ particles. In the three- and four-point cumulants that we are about to calculate, only two and three particles are free, respectively. Since these particles are indistinguishable, we can enumerate them with indices $(j_1,j_2) = (1,2)$ and $(j_1,j_2,j_3)=(1,2,3)$ and multiply the results with the number of ways to choose particle pairs and particle triples from an ensemble of $N$ particles.

\subsection{Three-point cumulant $G_{B\rho\rho}(11'{-1'})$}

We begin with the cumulants derived from the generating functional $Z_0^{(1)}[\tens L]$ from (\ref{eq:04-20}), which contains momentum correlations to linear order only. For $m = 3$, the one-particle response-field factor in (\ref{eq:04-10}) reduces to the single term
\begin{equation}
  b_{j_2'}(2') = -\ii\,g_{qp}(\tau_1,\tau_1')\,\vec k_1'\cdot\vec k_1\,
  \delta_{j_1j_2'}
\label{eq:04-44}
\end{equation}
because $\tau_1' = \tau_2'$ and therefore $g_{qp}(\tau_1',\tau_2') = g_{qp}(\tau_2',\tau_2') = 0$. Moreover, we have replaced $\vec k_2'$ by $-\vec k_1'$. Since the Kronecker symbol in the response-field factor couples the particles $j_1$ and $j_2'$, only two particle indices are free, which we set without loss of generality to $(j_1,j_1')=(1,2)$. The shift vectors $\bar L_{q_j}$ are then
\begin{equation}
  \bar L_{q_j} = -\left(\vec k_1-\vec k_1'\right)\delta_{j1}-\vec k_1'\delta_{j2}\;.
\label{eq:04-45}
\end{equation} 
For the two-point term (\ref{eq:04-21}), we can label the two points by $(j,k)=(1,2)$ and thus write
\begin{equation}
  \bar L_{q_1} = -\left(\vec k_1-\vec k_1'\right)\;,\quad
  \bar L_{q_2} = -\vec k_1'\;.
\label{eq:04-46}
\end{equation} 
We can stop here: The delta distribution in the two-point term in (\ref{eq:04-10}) shrinks to
\begin{equation}
  \delta_\mathrm{D}\left(\bar L_{q_1}+\bar L_{q_2}\right) = \delta_\mathrm{D}\left(\vec k_1\right)
\label{eq:04-47}
\end{equation}
and ensures this way that $\vec k_1 = 0$, which sets the response-field factor (\ref{eq:04-44}) to zero. We can thus conclude that $G_{B\rho\rho}(11'{-1'})$ cannot contribute at all to the one-point cumulant, hence
\begin{equation}
  \delta^{(1)}G_\rho(1) = 0
\label{eq:04-48}
\end{equation}
to first order in the interaction and to linear order in the momentum correlations: To this order, the interaction does not change the mean density.

For the two-point term (\ref{eq:04-22}) contributing to the quadratic momentum correlation, we can also set $(j,k)=(1,2)$ and arrived at the same conclusion: The delta distribution ensures $\vec k_1 = 0$ and thus sets the response to zero. The three-point term cannot contribute because $\bar L_{q_3} = 0$ according to (\ref{eq:04-45}), which implies $a_{23}^2 = 0$.

Of course, this is not surprising: No interaction can change the mean density in a canonical ensemble. It is merely reassuring to see why the individual contributions disappear formally.

\subsection{Four-point cumulant $G_{B\rho\rho\rho}(121'{-1'})$ from linear momentum correlations}

Turning to the effect of first-order interactions on the density power spectrum, we need to work out the four-point cumulant $G_{B\rho\rho\rho}(121'{-1'})$. The response-field factor is
\begin{align}
  b_{j_2'}(2') &=
  -\ii g_{qp}(\tau_1,\tau_1')\left(
    \vec k_1\cdot\vec k_1^{\,\prime}\,\delta_{j_1j_2'}+
    \vec k_2\cdot\vec k_1^{\,\prime}\,\delta_{j_2j_2'}
  \right)\;,
\label{eq:04-49}
\end{align}
setting $\vec k_2'=-\vec k_1'$ again. Other terms do not appear here because $g_{qp}(\tau_1',\tau_2') = 0 = g_{qp}(\tau_2',\tau_2')$. We shall further consider synchronous correlations only and thus set $\tau_1 = \tau_2$. Of the two terms remaining in (\ref{eq:04-49}), we now focus on the first, in which the Kronecker symbol ensures that $j_1 = j_2'$. The second term will then be obtained from the result by interchanging the indices $j_1$ and $j_2$ or, equivalently, the wave vectors $\vec k_1$ and $\vec k_2$.

Due to the coupling of two particles, three particles remain free, for which we choose the indices $(j_1,j_2,j_1')=(1,2,3)$ without loss of generality. The shift vectors are then
\begin{equation}
  \bar L_{q_j} = -\left(\vec k_1-\vec k_1'\right)\delta_{j1}-
  \vec k_2\delta_{j2}-\vec k_1'\delta_{j3}\;.
\label{eq:04-50}
\end{equation} 

The three particles need to be placed on three different positions to achieve the largest possible multiplicity. We choose three positions labelled by $(j,k,l)=(2,3,1)$, obtain the shift vectors
\begin{equation}
  \bar L_{q_j} = -\vec k_2\;,\quad
  \bar L_{q_k} = -\vec k_1'\;,\quad
  \bar L_{q_l} = -\left(\vec k_1-\vec k_1'\right)
\label{eq:04-51}
\end{equation}
from (\ref{eq:04-50}) and
\begin{equation}
  \bar Z_{23}^{(1)} =
  (2\pi)^3\delta_\mathrm{D}\left(\vec k_2+\vec k_1'\right)
  (2\pi)^3\delta_\mathrm{D}\left(\vec k_1-\vec k_1'\right)
  P_\delta\left(\vec k_2\right)A_{jk}^2\left(\vec k_2\right)
\label{eq:04-52}
\end{equation}
from (\ref{eq:04-21}). The second delta distribution arises from the factor $\mathcal{N}_{jk}'$. Since it ensures $\vec k_1'=\vec k_1$, it allows us to write (\ref{eq:04-52}) as
\begin{equation}
  \bar Z_{23}^{(1)} =
  (2\pi)^3\delta_\mathrm{D}\left(\vec k_1+\vec k_2\right)
  (2\pi)^3\delta_\mathrm{D}\left(\vec k_1-\vec k_1'\right)
  P_\delta\left(\vec k_1\right)A_{jk}^2\left(\vec k_1\right)\;,
\label{eq:04-53}
\end{equation}
where $A_{jk}(\vec k_1)$ simplifies to
\begin{equation}
  A_{jk}^2 = \frac{1}{2}\left(
    1+g_{qp}(\tau_1,0)g_{qp}(\tau_1',0)
  \right)+g_{qp}(\tau_1',0)
\label{eq:04-54}
\end{equation}
because $\vec k_1 = \vec k_1' = -\vec k_2$ due to the delta distributions. Permutations of $(j,k,l)$ with $l\ne1$, the factor $\mathcal{N}_{jk}'$ results in a delta distribution setting one individual wave vector to zero, which causes the result to vanish. The only other permutation leading to a non-vanishing result is $(j,k,l)=(3,2,1)$, for which
\begin{equation}
  A_{jk}^2 = \frac{1}{2}\left(
    1+g_{qp}(\tau_1,0)g_{qp}(\tau_1',0)
  \right)+g_{qp}(\tau_1,0)\;.
\label{eq:04-55}
\end{equation} 

After collecting results, the summation over particle indices multiplies the result by $N(N-1)(N-2)\approx N^3$, and the relevant two-particle contribution to the four-point density cumulant turns out to be
\begin{align}
  &G_{\rho\rho\rho\rho}(121'{-1'}) =
  \e^{-Q_D/2}\bar\rho^3(2\pi)^6\delta_\mathrm{D}\left(\vec k_1+\vec k_2\right)
  \delta_\mathrm{D}\left(\vec k_1-\vec k_1'\right) \nonumber\\ &\cdot
  \left(1+g_{qp}(\tau_1',0)\right)\left(1+g_{qp}(\tau_1,0)\right)
  P_\delta\left(\vec k_1\,\right)\;.
\label{eq:04-56}
\end{align}

Recall that this result was obtained assuming $j_1 = j_2'$. It is quite straightforward to see that the contribution for $j_2 = j_2'$ is identical, multiplying the cumulant by two. Thus, the four-point cumulant required for the first-order perturbation theory according to (\ref{eq:04-15}) is
\begin{align}
  &G_{B\rho\rho\rho}(121'{-1'}) =
  -2\ii\,\e^{-Q_D/2}\,g_{qp}(\tau_1,\tau_1') \nonumber\\
  &\times
  \left(1+g_{qp}(\tau_1',0)\right)\left(1+g_{qp}(\tau_1,0)\right)
  \nonumber\\ &\times \bar\rho^3
  (2\pi)^6\delta_\mathrm{D}\left(\vec k_1+\vec k_2\right)
  \delta_\mathrm{D}\left(\vec k_1-\vec k_1'\right)
  k_1^2\,P_\delta\left(\vec k_1\,\right)\;.
\label{eq:04-57}
\end{align}
With
\begin{equation}
  \bar{\tens L}_p^2 = \sum_{r,s=1}^m\vec K_r\cdot\vec K_s\,\delta_{j_rj_s}\;,
\label{eq:04-58}
\end{equation}
the damping term turns out to be
\begin{equation}
  Q_D = \frac{2\sigma_1^2}{3}\left(
    K_1^2-\vec K_1\cdot\vec K_1'+K_1'^2
  \right)\;.
\label{eq:04-59}
\end{equation}
According to (\ref{eq:04-15}), this implies the contribution
\begin{align}
\label{eq:04-60}
  &\delta^{(1)}G^{(1)}_{\rho\rho}(12) =
  -2(2\pi)^3\delta_\mathrm{D}\left(\vec k_1+\vec k_2\right)
  k_1^2\,P_\delta\left(\vec k_1\,\right) \\ &\cdot
  \int_0^{\tau_1}\d\tau_1'\,\hat v\left(\vec k_1\right)
  \e^{-Q_D/2}\,g_{qp}(\tau_1,\tau_1')\nonumber\\ &\times
  \left(1+g_{qp}(\tau_1',0)\right)\left(1+g_{qp}(\tau_1,0)\right)
  \nonumber
\end{align}
to the non-linear power spectrum, where the potential $\hat v(\vec k_1)$ was included in the time integral because its amplitude may depend on time, and the damping term $\e^{-Q_D/2}$ was included there because it does depend on time according to (\ref{eq:04-59}).

\subsection{Four-point cumulant $G_{B\rho\rho\rho}(121'{-1'})$ from quadratic momentum correlations}

We now turn to evaluating the contributions to the density power spectrum from quadratic initial momentum correlations, which are expressed by the free generating functional $\bar Z_0^{(2)}[\tens L]$ from (\ref{eq:04-20}). Since the response-field prefactor in (\ref{eq:04-49}) couples particle pairs, only three particle indices are free, which immediately implies that no four-point terms can contribute. The two- and three-point terms from (\ref{eq:04-22}) and (\ref{eq:04-23}) are thus the only ones to consider. Again, we label the particles by $(j_1,j_2,j_1')=(1,2,3)$ without loss of generality.

Regarding the three-point term $\bar Z_{jkkl}^{(2)}$, the position-index combination $(j,k,l) = (1,2,3)$ leads to
\begin{align}
  \bar Z_{1223}^{(2A)} &= \frac{\vec K_1\cdot\vec K_1'}{k_1'^2}
  \frac
   {\vec K_1\cdot(\vec k_1-\vec k_1')
    (\vec K_1-\vec K_1')\cdot(\vec k_1-\vec k_1')}
   {(\vec k_1-\vec k_1')^4} \nonumber\\ &\times
  P_\delta\left(\vec k_1-\vec k_1'\right)P_\delta\left(\vec k_1'\right)\;,
\label{eq:04-61}
\end{align}
the combination $(j,k,l) = (2,3,1)$ gives
\begin{align}
  \bar Z_{1332}^{(2B)} &= -\frac{\vec K_1\cdot\vec K_1'}{k_1^2}
  \frac
   {\vec K_1'\cdot(\vec k_1-\vec k_1')
    (\vec K_1-\vec K_1')\cdot(\vec k_1-\vec k_1')}
   {(\vec k_1-\vec k_1')^4} \nonumber\\ &\times
  P_\delta\left(\vec k_1\right)P_\delta\left(\vec k_1-\vec k_1'\right)\;,
\label{eq:04-62}
\end{align}
and the combination $(j,k,l)=(3,1,2)$ produces
\begin{equation}
  \bar Z_{3112}^{(2C)} = -\frac{\vec K_1\cdot(\vec K_1-\vec K_1')}{k_1^2}
  \frac{\vec K_1'\cdot(\vec K_1-\vec K_1')}{k_1'^2}\,
  P_\delta\left(\vec k_1\right)P_\delta\left(\vec k_1'\right)\;.
\label{eq:04-63}
\end{equation}

Finally, for the two-point term in (\ref{eq:04-22}) to contribute, the factor $\mathcal{N}_{jk}'$ returns a delta distribution for an individual wave number except for the particle-index combinations $(j,k,l)=(2,3,1)$ or $(3,2,1)$. For these,
\begin{align}
  \bar Z_{2323}^{(2D)} &=
  \delta_\mathrm{D}\left(\vec k_1-\vec k_1'\right)\,
  \int\frac{\d^3k}{(2\pi)^3}
  P_\delta\left(\vec k_1-\vec k\right)P_\delta\left(\vec k\right)
  \nonumber\\ &\times
  \frac{\vec K_1\cdot\vec k\,\vec K_1'\cdot\vec k}{k^4}
  \frac
   {\vec K_1\cdot(\vec k_1-\vec k)\vec K_1'\cdot(\vec k_1-\vec k)}
   {(\vec k_1-\vec k)^4}\;.
\label{eq:04-65}
\end{align}
For all terms in (\ref{eq:04-61}), (\ref{eq:04-62}), (\ref{eq:04-63}) and (\ref{eq:04-65}), the damping term agrees with (\ref{eq:04-59}).

The expressions (\ref{eq:04-61}), (\ref{eq:04-62}) and (\ref{eq:04-63}) each have the multiplicity $2^3=8$ due to the symmetry of the three-point term (\ref{eq:04-23}), while the expression (\ref{eq:04-65}) has the multiplicity $2^2=4$. Summing over all particle indices further multiplies the results by $N(N-1)(N-2) \approx N^3$. Taking the respective factors into account, we arrive at the relevant contribution
\begin{align}
  G_{\rho\rho\rho\rho}^{(2)}(121'{-1'}) = \bar\rho^3\e^{-Q_D/2}
  \left(
    \bar Z_{1223}^{(2A)}+\bar Z_{2331}^{(2B)}+\bar Z_{3112}^{(2C)}+
    \frac{\bar Z_{2323}^{(2D)}}{2}
  \right)
\label{eq:04-66a}
\end{align}
to the four-point density cumulant.

The contribution $G^{(2)}_{B\rho\rho\rho}(121'{-1'})$ of these terms to the cumulant $G_{B\rho\rho\rho}(121'{-1'})$ follows again by multiplying with the response-field factor (\ref{eq:04-49}), taking into account that both terms lead to same result. Thus,
\begin{equation}
  G^{(2)}_{B\rho\rho\rho}(121'{-1'}) = -2\ii g_{qp}(\tau_1,\tau_1')\,
  \vec k_1\cdot\vec k_1'\,G^{(2)}_{\rho\rho\rho\rho}(121'{-1'})\;.
\label{eq:04-67}
\end{equation}
Inserting this into (\ref{eq:04-15}), we find
\begin{align}
  \delta^{(1)}G_{\rho\rho}^{(2)}(12) &=
  -2\int_0^{\tau_1}\d\tau_1'
  g_{qp}(\tau_1,\tau_1')\nonumber\\ &\times
  \int\frac{\d^3\vec k_1'}{(2\pi)^3}
  \vec k_1\cdot\vec k_1'\,\hat v\left(\vec k_1'\right)
  G^{(2)}_{\rho\rho\rho\rho}(121'{-1'})\;.
\label{eq:04-69}
\end{align}

\section{First-order non-linear cosmic-density power spectrum}

\subsection{Initial, free evolution with the Zel'dovich propagator}

We have seen in \cite{2014arXiv1411.1153B} that the free non-equilibrium field theory for classical particles, beginning with an initially correlated point set in phase space, can naturally reproduce the well-known linear growth of the cosmic-density power spectrum. There, we found that the two-point density cumulant at the time $\tau$ is given by
\begin{equation}
  G_{\rho\rho}^{(1)}(12) = \bar\rho^2
  (2\pi)^3\delta_\mathrm{D}\left(\vec k_1+\vec k_2\right)
  \left(1+g_{qp}(\tau_1,0)\right)^2P_\delta\left(\vec k_1\right)
\label{eq:04-70}
\end{equation}
if we restrict the initial momentum correlations to their linear contribution. We ignore the damping factor $\exp(-Q_D/2)$ here for reasons detailed in \cite{2014arXiv1411.0805B}. It should in fact be neglected at linear order in the momentum correlations.

This result expresses the linear growth of the power spectrum if the propagator $g_{qp}(\tau,0) = \tau$ is used, which reflects the Zel'dovich approximation (see \cite{2014arXiv1411.0805B}). Since the cosmic time is chosen to be the linear growth factor of density fluctuations, $\tau = D_+-1$, the two-point synchronous density cumulant (\ref{eq:04-70}) recovers the linear growth of the power spectrum exactly.

Quadratic momentum correlations in the free theory produce a lowest-order deviation from the linear result (\ref{eq:04-70}), which was shown to be
\begin{align}
  &G_{\rho\rho}^{(2)}(12) = \e^{-Q_D/2}\frac{\bar\rho^2}{2}
  (2\pi)^3\delta_\mathrm{D}\left(\vec k_1+\vec k_2\right)
  g_{qp}^4(\tau_1,0) \nonumber\\ &
  \times\int\frac{\d^3k}{(2\pi)^3}
  P_\delta\left(\vec k\,\right)P_\delta\left(\vec k-\vec k_1\right)
  \left(\frac{\vec k_1\cdot\vec k}{k^2}\right)^2
  \left(
    \frac{\vec k_1\cdot(\vec k-\vec k_1)}{(\vec k-\vec k_1)^2}
  \right)^2
\label{eq:04-71}
\end{align}
in \cite{2014arXiv1411.0805B}. There, we also explained that the damping factor should be approximated to first order by
\begin{equation}
  \e^{-Q_D/2} = \frac{1}{\e^{Q_D/2}} \approx \frac{1}{1+Q_D/2}
\label{eq:04-72}
\end{equation}
if quadratic momentum correlations are included.

While the contributions (\ref{eq:04-71}) and (\ref{eq:04-72}) to the two-point density cumulant thus reproduce the linear growth of the power spectrum and add a first non-linear term, it should be noted that this growth and this onset of the non-linear evolution occur in the \emph{free} theory already because we have inserted the Zel'dovich propagator there. As described in \cite{2014arXiv1411.0805B} and many other studies, the Zel'dovich approximation is remarkably successful in cosmology because it captures a substantial fraction of the interaction potential between the particles. It is this part of the interaction, contained in the free Zel'dovich propagator, which gives rise to the structure growth expressed by (\ref{eq:04-71}) and (\ref{eq:04-72}) even in our free theory.

\subsection{Subsequent evolution with the Hamiltonian propagator}

The further contributions $\delta^{(1)}G_{\rho\rho}^{(1)}(12)$ given in (\ref{eq:04-62}) and $\delta^{(1)}G_{\rho\rho}^{(2)}(12)$ from (\ref{eq:04-69}) above are of a different nature. There, the interaction potential is explicitly taken into account, if only at linear order. It is conceptually opaque to consistently combine both approaches, i.e.\ the evolution with the Zel'dovich propagator which implicitly includes part of the interaction, and the evolution with an explicit interaction potential, from which the contribution contained in the Zel'dovich propagator would have to be removed. The main reason is, as detailed in \cite{2014arXiv1411.0805B}, that the Zel'dovich approximation combines different times in the description of particle trajectories, namely the initial time when the Zel'dovich velocity potential needs to be evaluated, with the final time of the particle position along its trajectory.

We avoid this difficulty in the following way, which is closely modelled on the procedure commonly followed in numerical simulations. We use the free theory with the Zel'dovich propagator from an early, but otherwise arbitrary initial time $\tau_\mathrm{i} = 0$ to a time $\tau_*$ yet to be determined. At this point in time, we switch to the explicit, perturbative account of the interaction potential. This implies that we shall replace the Zel'dovich propagator by the propagator
\begin{equation}
  g_{qp}(\tau,\tau') = \int_{\tau'}^\tau\frac{\d\bar\tau}{g(\bar\tau)}\;,
\label{eq:04-73}
\end{equation}
for Hamiltonian particles in an expanding space-time, with
\begin{equation}
  g(\tau) = a^2D_+(a)f\frac{H}{H_\mathrm{i}}\;,
\label{eq:04-74}
\end{equation}
where $H$ is the Hubble function and $H_\mathrm{i}$ its value at the initial time, and
\begin{equation}
  f := \frac{\d\ln D_+}{\d\ln a}\;.
\label{eq:04-75}
\end{equation}

The time $\tau_*$ is uniquely and fully specified in the following way. We have three terms which are proportional to the initial density power spectrum $P_\delta^\mathrm{(i)}(k)$ with its shape unchanged. The first is the term $G^{(1)}_{\rho\rho}(12)$ from (\ref{eq:04-70}) caused by the free evolution of linear momentum correlations, which occurs twice: one time evolved with the Zel'dovich propagator from the initial time to $\tau_*$, and the second time evolved with the Hamiltonian propagator from $\tau_*$ to $\tau_1$. The second term is $\delta^{(1)}G_{\rho\rho}^{(1)}(12)$ from (\ref{eq:04-60}), which occurs only between $\tau_*$ and $\tau_1$.

The term $G_{\rho\rho}^{(1)}(12)$ implies the power spectrum
\begin{equation}
  P_\delta^{(1)}(k,\tau_*) = D_+^2(\tau_*)P_\delta^\mathrm{(i)}(k)
\label{eq:04-76}
\end{equation}
at $\tau_*$, starting from the primordial density power spectrum $P_\delta^\mathrm{(i)}(k)$. The further evolution from $\tau_*$ to $\tau_1$ creates the power-spectrum contribution
\begin{equation}
  P_\delta^{(1)}(k,\tau_1) = \left(1+g_{qp}(\tau_1,\tau_*)\right)^2P_\delta^{(1)}(k,\tau_*)\;.
\label{eq:04-76a}
\end{equation}
The term $\delta^{(1)}G_{\rho\rho}^{(1)}(12)$ adds
\begin{equation}
  \delta^{(1)}P_\delta^{(1)} = \Delta^2(\tau,\tau_*)P_\delta^{(1)}(k,\tau_*)
\label{eq:04-77}
\end{equation}
to the power spectrum, with
\begin{align}
  \Delta^2(\tau,\tau_*) &:= 3\int_{\tau_*}^\tau\d\tau'
  \frac{ag_{qp}(\tau,\tau')}{g(\tau')}\nonumber\\ &\times
  \left(1+g_{qp}(\tau_1',0)\right)\left(1+g_{qp}(\tau_1,0)\right)\;,
\label{eq:04-78}
\end{align}
which combines the explicit time dependence shown in (\ref{eq:04-60}) with the time dependence of the Poisson equation (\ref{eq:04-38}). The condition defining $\tau_*$ is thus
\begin{equation}
  D_+^2(\tau_1)P_\delta^\mathrm{(i)}(k) = \left\{
    1+
    \left(1+g_{qp}(\tau_1,\tau_*)\right)^2+
    \Delta^2(\tau)
  \right\}P_\delta^{(1)}(k,\tau_*)\;.
\label{eq:04-79}
\end{equation}
Substituting $P_\delta^{(1)}(k,\tau_*)$ from (\ref{eq:04-76}) gives the implicit equation
\begin{equation}
  \frac{D_+^2(\tau_1)}{D_+^2(\tau_*)} = 1+
  \left(1+g_{qp}(\tau_1,\tau_*)\right)^2+\Delta^2(\tau,\tau_*)
\label{eq:04-80}
\end{equation}
for $\tau_*$, which is easily solved numerically. Typically, we find values of $\tau_* \approx 100$ for Friedmann cosmologies with CDM power spectra normalised to $0.8\le\sigma_8\le1$.

\subsection{Results}

\begin{figure}[ht]
  \includegraphics[width=\hsize]{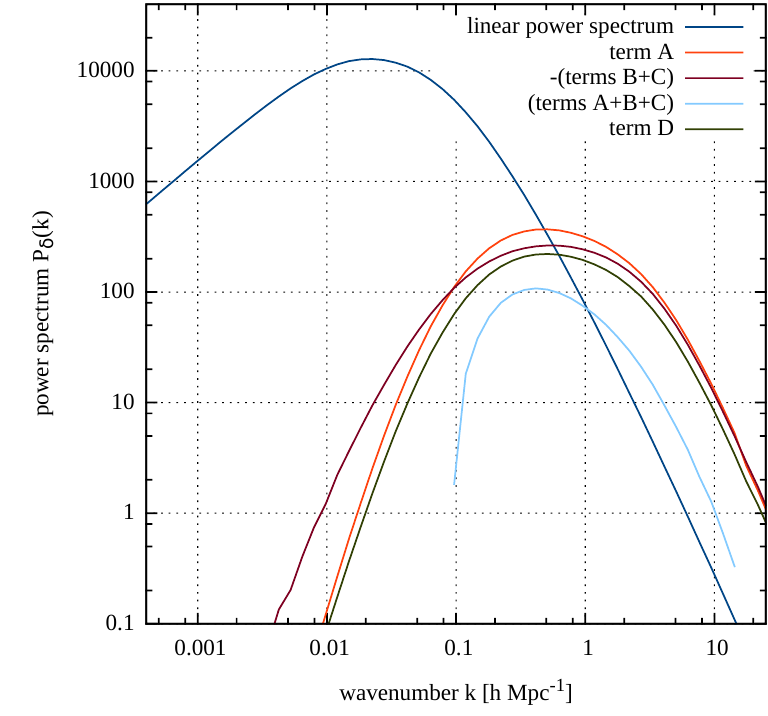}
\caption{The terms contributing to $\delta^{(1)}G^{(2)}_{\rho\rho}(12)$ are shown, labelled by $A\ldots D$ as in (\ref{eq:04-63}) to (\ref{eq:04-67}). The blue curve with the largest amplitude is the CDM power spectrum linearly evolved to $z=0$, for comparison. The damping term was taken into account to first order, as shown in (\ref{eq:04-72}). The underlying cosmological model is a standard $\Lambda$CDM model with $\Omega_\mathrm{m0} = 0.3$ and $\Omega_\mathrm{\Lambda0} = 0.7$, normalised to $\sigma_8 = 0.8$.}
\label{fig:04-1}
\end{figure}

\begin{figure}[ht]
  \includegraphics[width=\hsize]{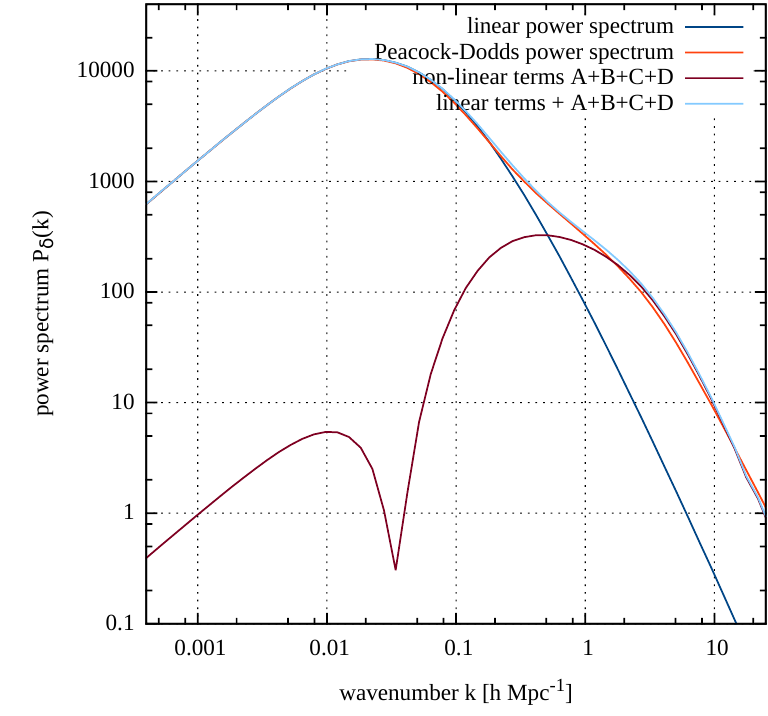}
\caption{The non-linear contributions to the power spectrum are shown together with the linearly evolved power spectrum, and the sum of the linear and non-linear terms is compared to the non-linear power spectrum according to the recipe derived from numerical simulations by \cite{1996MNRAS.280L..19P}. Cosmological parameters were chosen as for the curves shown in Fig.~\ref{fig:04-1}. Our analytic result agrees well with the non-linear power spectrum according to Peacock and Dodds.}
\label{fig:04-2}
\end{figure}

With $\tau_*$ fixed, we can proceed to calculating the non-linear cosmic density power spectrum to first order in the particle interaction. The interaction terms contributing to the non-linear power spectrum are shown individually in Fig.~\ref{fig:04-1}. Fig.~\ref{fig:04-2} shows the sum of the linear and non-linear contributions compared to the non-linear power spectrum derived by \cite{1996MNRAS.280L..19P} from numerical simulations.

As Fig.~\ref{fig:04-2} shows, the agreement between our analytic results and the non-linear power spectrum extracted from numerical simulations is quite good.

These results seem to indicate that our statistical non-equilibrium field theory for classical microscopic particles allows us to calculate the statistics of the non-linear evolution of cosmic density fluctuations quite accurately even at low orders of the interaction potential. Quantitatively, the results shown in Figs.~\ref{fig:04-1} and \ref{fig:04-2} should still be taken with caution, however. As we have described earlier in this section, we can achieve these results by switching from a Zel'dovich evolution phase and a subsequent Hamiltonian phase at a cosmic time chosen such as to arrive at the known amplitude of the linear power spectrum today. While this may be seen as equivalent to the approach often followed in numerical simulations, it would of course be more satisfactory to have one prescription, and thus one propagator, valid for the entire evolution. The improvement of the Zel'dovich approximation derived in \cite{2014arXiv1411.0805B} may offer a way towards this, which we are currently studying.

Furthermore, the damping factor approximated in (\ref{eq:04-72}) plays a crucial role here. So far, we have set the damping length to its comoving value at $\tau_*$ and neglected any time evolution. While this treatment should be appropriate in the context of a first-order calculation, the damping factor needs to be considered with more care in more advanced calculations aiming at a detailed, quantitative comparison with numerical simulations. Then, terms of second order in the interaction potential will also be included. We are currently extending the theory into this direction, which is naturally quite involved.

One aspect of our results that should perhaps be emphasised is that all terms except term $C$ from (\ref{eq:04-63}) contain convolutions of the initial power spectrum with itself. This is natural for a one-loop perturbative calculation. The convolved spectra, when multiplied with the first-order approximation of the damping factor, ensures the correct asymptotic behaviour of the power spectrum for large wave numbers. We may expect that, as both the loop order of the perturbative calculation and the approximation order for the damping factor are increased, the asymptotic behaviour of the power spectrum is preserved. This is one of many properties of the theory to be studied.

\section{Conclusions}

Starting with a non-equilibrium, statistical field theory for microscopic, classical degrees of freedom, we have derived first-order perturbative corrections to the density power spectrum of a canonical particle ensemble initially correlated in phase space. While our main target is cosmological structure formation, the theory is generally valid for classical $N$-particle ensembles with arbitrary initial conditions and interaction potentials.

Specialising our results to initial conditions and propagators appropriate for cosmology, we have shown that the perturbative terms of first order in the interaction are reproducing the non-linear evolution of the density power spectrum observed in numerical simulations quite well. Our calculation extends to redshift zero and to arbitrary wave numbers. It has no free parameters once the power spectrum of the initial phase-space particle distribution is fixed and normalised. The form of the non-linear terms and the inevitable damping factor suggest that the expected asymptotic behaviour of the non-linear power spectrum for large wave numbers will be retained in higher-order calculations.

The main difference to ordinary, Eulerian or Lagrangian perturbation theory of cosmic-structure evolution is that we do not require, solve or perturb a dynamical equation for the cosmic density. Rather, we study the statistical evolution of a particle ensemble in phase space, weakly perturbing their trajectories, and read any collective information such as the density of the evolved phase-space distribution when needed. Since even small perturbations of trajectories can lead to large increases in density, our approach is able to extend into the regime of highly non-linear density perturbations even at low perturbative orders. It also appears crucial to keep the complete phase-space information of the particles because this allows us to use the Hamiltonian equations of motion with their simple structure and their equally simple Green's function.

On the way to our non-linear results, we had to switch from Zel'dovich to Hamiltonian propagators at a time set by the theory itself. This approach needs to be improved by propagators suitable for all time. Moreover, the detailed treatment of the damping factor and its time evolution require further study. Nonetheless, the first-order results we have achieved here seem to show that the theory developed here will be capable of describing the fully non-linear, statistical evolution of classical particle ensembles in general, and of cosmological structures in particular.

\begin{acknowledgments}
We wish to thank Luca Amendola, J\"urgen Berges, Marc Kamionkowski, Brice M\'enard, Adi Nusser, Manfred Salmhofer, Bj\"orn Sch\"afer, Naoshi Sugiyama, Alex Szalay, Christof Wetterich and Saleem Zaroubi for inspiring and helpful discussions. This work was supported in part by the Transregional Collaborative Research Centre TR~33, ``The Dark Universe'', of the German Science Foundation. The generous support by J\"urgen Berges and the Institute for Theoretical Physics at Heidelberg University was essential for this study.
\end{acknowledgments}

\bibliography{../bibliography/main}

\end{document}